\documentclass[12pt,letterpaper]{article}
\usepackage[utf8]{inputenc}
\usepackage[english]{babel}
\usepackage[vmargin=0.6in,hmargin={1in,1in},headsep=0.3in,headheight=1in]{geometry}
\usepackage{subfiles}
\usepackage{tikz}
\usepackage{fix-cm}
\usepackage{enumitem}
\usepackage{multicol}
\usepackage{amsmath, amsthm, amssymb, amsfonts,mathrsfs}
\usepackage{paracol}
\usepackage{graphicx}

\allowdisplaybreaks
\newcounter{taggedEquations}
\let\OldTag\tag
\renewcommand*{\tag}[1]{\stepcounter{taggedEquations}\OldTag{#1}}
\usepackage{array}
\usepackage{tabularx}

\usepackage{relsize}
\usepackage{xcolor}
\usepackage{lastpage}
\usepackage{blindtext}
\usepackage[numbers,sort&compress]{natbib}
\usepackage[linktocpage=true]{hyperref}
\hypersetup{
	colorlinks,
	citecolor=blue,
	filecolor=black,
	linkcolor=blue,
	urlcolor=blue
}
\usepackage{url}

\usepackage{graphicx}
\usepackage{caption}
\usepackage{subcaption}
\usepackage{wrapfig}
\usepackage{indentfirst}
\usepackage{hyperref}
\usepackage{soul}
\usepackage{bm}

\usepackage{empheq}
\usepackage{authblk}
\setlist{parsep=0pt,listparindent=\parindent}
\setlist[itemize]{noitemsep, topsep=0pt}
\setlist[enumerate]{noitemsep, topsep=0pt}
\setlist{parsep=0pt,listparindent=\parindent}

\stepcounter{footnote}
\graphicspath{{Images/}{../Images/}}
\usepackage{graphicx}
\graphicspath{ {./images/} }
\usepackage{pgfplots}
\usepackage{extarrows}
\usepackage{mathtools}
\usepackage{tikz}
\usetikzlibrary{decorations.pathmorphing,patterns}
\usetikzlibrary{decorations.pathreplacing,angles,quotes}
\usetikzlibrary{patterns}
\usetikzlibrary{shapes.arrows, fadings}
\usepackage{xfrac} 
\usepackage{relsize}
\usepackage[most]{tcolorbox}
\tikzfading[name=fade down,
  top color=transparent!0, bottom color=transparent!100]
\tikzfading[name=fade up,
  bottom color=transparent!0, top color=transparent!100]

\usepackage{pgfplots}
\pgfplotsset{compat=1.18}
\usetikzlibrary{external}

\usepackage{physics}

\usepackage{enumitem}

\def\beq{\begin{eqnarray}}  
\def\eeq{\end{eqnarray}}

\begin{document}

\title{\Large\textbf{Killing Invariants: An approach to the sub-classification of geometries with symmetry}}
\author[1,2]{\large{Brown, C. }}
\author[1,2]{\large{Gorban, M.}}
\author[1,2]{\large{Julius, W.}}
\author[1,2]{\large{ Radhakrishnan, R.}}
\author[1,2]{\large{Cleaver, G. }}
\author[3]{\large{McNutt, D., }}

\affil[1]{\emph{Early Universe, Cosmology and Strings (EUCOS) Group, Center for Astrophysics, Space Physics and Engineering Research (CASPER)}}
\vspace{1 cm}
\affil[2]{\emph{Department of Physics, Baylor University,  Waco, TX 76798, USA} }
\vspace{1 cm}
\affil[3]{\emph{UiT: The Arctic University of Norway, Department of Mathematics and Statistics, 9019, Tromsø, Norway, }}
\vspace{0.25cm}

\date{\today}  
\maketitle  
\begin{abstract}   

In principle, the local classification of spacetimes is always possible using the Cartan-Karlhede algorithm. However, in practice, the process of determining equivalence of two spacetimes relies on determining if a set of equations has a solution. Depending on the form of the equations this may be undecideable. Furthemore, if a solution does exist the equations may still be unsolvable in some way. In the case that spacetimes admit Killing vector fields with non-trivial orbits, we propose a new set of invariant quantities, called Killing invariants. These invariants will allow for the sub-classification of spacetimes admitting the same group of symmetries and will, in principle, be substantially less complicated than other sets of invariants. We apply this approach to the class of static spherically symmetric geometries as an illustrative example. 

\end{abstract} 

\par

\pagebreak

\section{Introduction}\label{section:intro}

The Cartan-Karlhede (CK) algorithm can always be applied to locally classify Lorentzian geometries (or spacetimes)  \cite{karlhede1980review}. It has been used to distinguish solutions within General Relativity (GR) \cite{kramer} and alternative metric-based gravity theories in four dimensions (4D) and higher dimensions \cite{mcnutt2017cartan}. In principle, the algorithm provides a method to determine when two metrics are equivalent; that is, when a coordinate transformation can be used to transform one metric into the other. However, in practice, the algorithm is primarily used to prove when spacetimes are inequivalent. This is due to the final step used to compare spacetimes, where one must equate the corresponding sets of Cartan invariants of each metric and solve the resulting equations.

This step is further complicated by the fact that it may not be possible to show a solution exists. That is, even if a solution exists, it may be impossible to prove that a given set of equations has a solution. Thus, the set of equations would be said to be undecidable. To illustrate this we will briefly discuss two common scenarios where the system of equations admit rational functions or transcendental functions in their expressions. In the first example, the equations are constructed from rational functions. Using Gr{\"o}bner bases it is possible to prove that a solution exists. However, if the rational functions are too large then the system of equations will be practically unsolvable due to limitations of available computation. In the second example, whether a set of equations is decidable will depend on the chosen transcendental functions. However, if the choice of functions is restricted to the sine function and the modulus function, then it has been shown that there are sets of equations with these functions that are undecidable \cite{richardson1969some}.

An excellent example of this difficulty appears within the class of stationary axisymmetric spacetimes with two commuting Killing vector fields \cite{kramer}. If a given spacetime admits an irreducible rank 2 Killing tensor field, then the metrics within this class of spacetimes  can be expressed in Boyer-Lindquist form  \cite{eichhorn2021locality}. Alternatively, if a metric can be expressed in Boyer-Lindquist form then the irreducible rank 2 Killing tensor has a canonical form in terms of the metric functions \cite[Appendix A.5]{eichhorn2021locality}. This is remarkable since, in general, it is difficult to determine the existence of an irreducible rank 2 Killing tensor \cite{kruglikov2022killing}. 

Given a stationary axisymmetric metric in an arbitrary coordinate system and one in Boyer-Lindquist coordinates, it is then possible to determine if they are equivalent. Necessary conditions for equivalence can be determined by comparing the discrete invariants generated in the algorithm, namely the number of functionally independent invariants and the dimension of a subgroup of the Lorentz frame transformations determined at each iteration of the algorithm. If two metrics share the same discrete invariants, then the final step must be initiated to show equivalence. while some of the metrics in this clas, such as the Kerr-like metrics,  will have simple coordinate expressions for their Cartan invariants. There will also be metrics where their Cartan invariants have complicated coordinate expressions, leading to potentially unsolvable equations in the comparison steps of the CK algorithm. 

This problem is of particular importance when exploring the class of axisymmetric and stationary vacuum solutions in GR beyond the Kerr solution. While there are uniqueness  theorems and classification results for the Kerr black hole \cite{robinson1975uniqueness, mars2000uniqueness, ferrando2009intrinsic}, the investigation of other stationary axisymmetric solutions within GR or alternative gravity theories \cite{Owen:2021eez, eichhorn2021locality} will require practical tools to classify solutions. The potential issue of complicated coordinate expressions for curvature invariants can be alleviated by introducing a new set of invariants for a spacetime, whose coordinate expressions are expected to be smaller than the curvature invariants. Alternative invariants have been proposed to classify Lorentzian geometries, either by studying the kinematic properties of a chosen frame basis for the spacetimes 
\cite{Kitamura1978} or by working with differential invariants associated with the jet bundle. This latter set of invariants has been used to classify the class of metrics admitting two commuting Killing vector fields  \cite{marvan2008local, ferraioli2020equivalence} as well as other metrics admitting larger Killing algebras, including static spherical symmetric metrics \cite{mcnutt2024detecting}. 

In this paper we will introduce a new set of invariants by considering geometric quantities generated from Killing vector fields in the spacetime. If a spacetime admits Killing vector fields, these are certainly invariantly defined up to scaling by constant values and linear combinations. A natural question to ask is whether or not scalar invariants can be constructed using them. However, before doing so we will review some applications of Killing vector fields in GR and other gravity theories.

\subsection{Applications of Killing vector fields }

Geometrically, the use of Killing vector fields as a tool to generate solutions is well-known \cite{schwarzschild1916gravitationsfeld,cahen1968lorentzian,kramer}. However, the perspective that Killing vector fields have some relevance to physics is less obvious. The existence of symmetries in a spacetime implies the existence of conserved quantities when studying geodesics \cite{Wald:1984rg}. More generally, it has been argued that the kinematics of spacelike or timelike geodesics that lie exclusively in the orbits of the symmetry group give rise to physically meaningful invariants \cite{ greenberg1970propagation, chinea1992differential, fernandez1999exterior}.

Fayos and Sopuerta \cite{fayos1999papapetrou, fayos2001general} detail a method of studying the local structure of spacetime by using the Papapetrou field, the exterior derivative of the Killing vector field in their definition, as the main algebraic object. They emphasize the use of the Killing vector field as a vector potential solution of Maxwell's equations and the Papapetrou field therefore being seen as an electromagnetic field tensor. Through the integrability conditions of the Killing vector fields, written in terms of the Papapetrou fields, they show that the Weyl tensor in 4D can be completely expressed in terms of the Killing vector field, the Papapetrou field, and the energy-momentum tensor. In addition, by algebraically classifying Papapetrou forms, they determine conditions for when the principal directions of a Papapetrou field aligned with the principal directions of the Weyl tensor and hence determine the Petrov type of the spacetime \cite{fayos1999papapetrou,fayos2001general,fayos2002consequences}. It is also shown that the Newman-Penrose formalism can be adapted into a form that places the Papapetrou field into a canonical form. These results on Papapetrou fields can also be extended by adding new variables derived from the Killing vector fields. Through this formalism, the Weyl tensor components can be algebraically expressed in terms of the matter field, the spin-coefficients, the eigenvalues of the Papapetrou field, the Killing vector field variables and the derivatives of a complex scalar   \cite{fayos2002consequences}. These results indicate that the study of Killing vector fields in general relativity can give meaningful information about the physical interpretation of solutions.

The use of invariant quantities derived from Killing invariants is not restricted to 4D GR or its modifications. For example, the classification of five-dimensional supersymmetric black holes with a single axial symmetry was accomplished using the norm of a Killing vector field as a global invariant for the solutions \cite{katona2023classification, katona2023supersymmetric}. In gravitational theories involving torsion, such as Einstein-Cartan or teleparallel gravity, a frame formalism can be adapted to the group of symmetries in order to derive the most general family of solutions that admit that symmetry group. These invariants also provide a direct way to sub-classify these families of solutions \cite{ mcnutt2023frame}.

\section{Killing invariants}

We will adopt a frame approach similar to the CK algorithm, although the algorithm will also share features of Cartan's method of moving frames \cite{olver1995equivalence}. In Cartan's approach, the parameters of the frame transformation group are first fixed by considering the exterior derivative of the coframe basis. If some parameters have not be fixed, the exterior derivatives of the components of the resulting two-forms are used to fix them. This process can continue iteratively by taking the exterior derivative of the components of the  resulting one-forms, until no further parameters can be fixed. If some parameters can never be fixed, this indicates the existence of some linear isotropy group.   

By using the fact that the Killing vector fields of a spacetime are invariantly defined, we will employ them to fix the frame in a coordinate independent manner by treating the Killing one-forms relative to a chosen coframe as an exterior differential system \cite{chinea1988symmetries, estabrook1996moving, papadopoulos2012locally}. This approach will take a given Lie algebra of Killing vector fields and generate a representative frame basis for the manifold. We note that while the stationary axisymmetric spacetimes have been discussed in order to motivate this algorithm, we will instead focus on the class of spherically symmetric spacetimes in order to introduce our formalism, and in a future paper we will apply this approach to stationary axisymmetric geometries.

As a partial answer to the question of scaling Killing vector fields by a constant value, we can employ the CK algorithm \cite[Chapter 9]{kramer} to determine an invariantly defined frame, up to freedom in the linear isotropy group. We will call this frame a {\it CK frame} and write this as $\{{\bf e}_a\}$ and the dual coframe as $\{ \theta^a\}$. In the case of non-trivial isotropy, the pullback of an arbitrary change of coordinates, i.e., an isometry, $\Phi: M \to M$, gives 
\beq
    \Phi_* \theta^a = \Lambda^a_{~b} \theta^b, 
\eeq
\noindent where $\Lambda^a_{~b}$ lies in the linear isotropy group $H_p(M)$.  If $\Phi_\tau$ is the flow of a Killing vector field, ${\bf X}$, at a point $p \in  M$, then taking the limit of $\tau\to 0$ we find 
\beq 
    \mathcal{L}_{{\bf X}} \theta^a = \lambda^a_{~b} \theta^b, 
\eeq

\noindent where $\lambda^a_{~b}$ lies within the Lie algebra of the linear isotropy group. For a given CK frame, we can solve the resulting frame based Killing equations to determine the form of the Killing vector fields \cite{karlhede1982determining, mcnutt2023frame}. In doing so, we can write any Killing vector field, ${\bf X}$, of a given spacetime in terms a CK frame where the coefficients, $X^a$, are invariant quantities, 
\beq 
    {\bf X} = X^a {\bf e}_a.
\eeq
\noindent The Killing vector fields are solutions of a system of linear first order partial differential equation called the Killing equations. It is useful to determine a basis of solutions. Relative to the CK frame, we can determine the set of Killing vector fields in a coordinate independent manner by identifying a linear independent basis of solutions to the Killing equations. We can effectively ignore the problem of scaling the Killing vector fields so long as an invariantly defined frame has been constructed and we are only comparing functional relationships between the resulting invariants. 

The CK frame does not resolve the ambiguity involved in constant linear combinations of Killing vector fields. For example, in the class of stationary axisymmetric spacetimes, we can apply a $GL(2, \mathbb{R})$ transformation to produce new timelike and spacelike Killing vector fields which cannot be distinguished from the original pair. This is not unexpected nor unwelcome, it is a natural byproduct of the symmetry of the spacetime. We note that the existence of a non-trivial Lie algebra can restrict the freedom of linear combinations by imposing a canonical form for the structure coefficients. 

\subsection{Killing invariants algorithm}

We will consider the case of an $N$-dimensional manifold. The procedure begins with the assumption that a basis for the set of all Killing vector fields, $\{ {\bf X}_A \}$, $A,B,E = 1,\ldots, m$ with $m \leq \frac{N(N+1)}{2}$  is known in the coordinate basis. This is a strong requirement, as one must necessarily solve the associated system of first order linear partial differential equations. While this is guaranteed to exist, it is a non-trivial calculation.  We will denote the Killing one-forms as ${\bf F}^A = g_{ab} X^a_{~A} \theta^b$. The structure coefficients $C^A_{~BE}$ and inner-products of the Killing vector fields, 
\beq  N_{AB} = g_{ab} X^{\,a}_{~A} X^{\,b}_{~B},  \eeq
are then computed to determine where the vector fields are timelike, spacelike, or null. We will call the inner-product of two copies of a vector field the norm.  We then determine if there are points in the manifold where the norm of the Killing vector field vanishes, $N_{AA} = 0 $, and $ X^a_{~A} = 0$. If this occurs for a given Killing vector it lies in the isotropy group. This motivates the assumption of regularity in an open region due to the potential change in the dimension of the isotropy group in spacetime. An example of this would be in the Szekeres Swiss Cheese models where spherically symmetric metrics are mached to quasi-spherical Szekeres metrics \cite{bolejko2010szekeres}. We are motivated in local invariants that characterize the geometry at an arbitrary point, and so we will restrict to regions where the dimension of the isotropy group is constant.

If a given Killing vector is not a member of the isotropy group it is instead a translational Killing vector, which we  collect and form the set  $\{ {\bf {\bf X}_I}\}$, $I, J, K = 1, \ldots m'$ where $m' \leq r$ and $r \leq N$ is the dimension of the orbits of the symmetry group. For example, in the case of a static and spherical symmetric spacetime in N dimensions, the orbits of the symmetry group is $m' = N-1$ whereas the number of translational Killing vector fields is at most $2$ in a coordinate system where one spatial rotation has been rectified. 

Taking the perspective that the $N_{IJ}$ have geometrical and physical significance, the vanishing of the norm, $N_{II}$, indicates where the translational Killing vector fields become null. For example, in the exterior of the ergosphere of the Kerr black hole, the timelike Killing vector field, $\partial_t$, becomes null at any point in the ergosphere. On surfaces where some $N_{II}$ vanish, it is not possible to construct an orthonormal frame adapted to the desired Killing vector field since it is now a null vector field. Thus, we must restrict the neighbourhood appropriately to work with an orthonormal frame. In the context of classification, this is not an obstacle and potentially provides a helpful diagnostic to check the inequivalence of two solutions.

In an open region where all $N_{II}$ are non-zero, the Gram-Schmidt procedure constructs an orthonormal frame adapted to the translational Killing vector fields. As an initial coframe, one could use the simplest coframe arising from the metric or the CK coframe. The result of this process is a coframe, $\{ \theta^a \}$, and a restricted subgroup of Lorentz frame transformations $H_{-1}$. This subgroup is related to the free parameters arising from the Gram-Schmidt procedure when there are less than $N$ translational Killing vector fields. We will denote the dual frame basis as $\{ \mathbf{e}_a \}$. Relative to this frame, the components of the translational Killing vector fields will be written as rational expressions of the $N_{IJ}$.

The algorithm to construct invariant quantities from a set of Killing vector fields is then:
\begin{enumerate}
\item Denote the set $\mathcal{K}_0 =  \{ N_{IJ} \} $ for $I, J = 1, \ldots, m'$ as the zeroth-order Killing invariants.
\item Set the order of differentiation q to one. 
\item Calculate the exterior derivatives of the Killing one-forms, $d{\bf F}^I = F^I_{~ab} \theta^a \wedge \theta^b$ where $F^I_{~ab} = e^{}_{[a}(F^I_{~b]})$ where $e_{a_i}$ acts as a frame derivative operator on a scalar.
\item Find the canonical form for each of the $d{\bf F}^I$.
\item Fix the frame parameters to realize the canonical forms of some of the $d{\bf F}^I$, and note the residual frame freedom, which we will call the linear isotropy group, $\bar{H}_0$. 
\vspace{2 mm}

\noindent The components of $d{\bf F}^I$ relative to this frame are called first-order Killing invariants, denoted as an ordered list \beq 
    \mathcal{K}_1 = \mathcal{K}_0 \cup \{ F^I_{~ab} \}, \quad\quad 0 \leq a<b \leq N. 
\eeq 
Similarly, we will introduce the ordered list $\mathcal{K}_q$ as the collection of all Killing invariants up to $q^{th}$-order, 

\beq 
    \mathcal{K}_q =  \mathcal{K}_{q-1} \cup \{ F^I_{~b a_1 a_2 \ldots a_q} \},
\eeq

\noindent where $ F^I_{~b a_1 a_2 \ldots a_q} = - e_{a_q} \ldots e_{a_2} e^{}_{[a_1}(F^I_{~b]})$.
\vspace{2 mm}

\item For $q\geq 1$ calculate the exterior derivatives of the elements in $\mathcal{K}_q$. 
\item Fix the frame as far as possible using the one-form, $d k$ with $k \in \mathcal{K}_q$, and note the residual frame freedom, which we will call the linear isotropy group, $\bar{H}_q$. 

\item If the linear isotropy group is the same as the previous iteration, let $p = q$ and stop; if they differ, or if $q=0$, increment $q$ by 1 and go to step 5.  
\end{enumerate}

\noindent Upon completing the algorithm we will denote $\mathcal{K} = \mathcal{K}_{p}$ as the set of all Killing invariants. Any spacetime admitting a symmetry group can then be characterized locally by the canonical form used and the successive isotropy groups. The Killing invariants in $\mathcal{K}$ can then be used to characterize the geometry. 

In principle, one could build scalar polynomial invariants in terms of the Killing invariants by taking contractions of even rank tensor built from tensor products of the metric along with $F^I, dF^I$ and the exterior derivatives of elements in $\mathcal{K}_q$, $q>1$. This has the advantage of side-stepping the choice of frame at the potential cost of more complicated invariants. 

To compare two spacetimes, $(M,{\bf g})$ and $(M', {\bf g'})$, we can use the descending chain of linear isotropy groups at each order to determine inequivalence. We note that comparing their respective sequences is not sufficient to determine equivalence. If the two sequences agree so that the spacetimes share similar canonical forms, it is necessary to compare the set of equations obtained by equating each entry in the ordered list $\mathcal{K}_{p}$ with the corresponding entry in $\mathcal{K}'_{p}$. If this set of equations has a solution, then the two spacetimes are equivalent. We note that this final step is not algorithmic and is still potentially undecidable if a solution exists.

\section{Static spherically symmetric metrics} \label{section:SSSmetrics}

To illustrate how the algorithm can be implemented when non-trivial isotropy is present, we consider the class of N+2-dimensional spherically symmetric metrics relative to the spherical coordinate system $(t,r,\phi_!, \ldots \phi_N)$,
\beq 
    ds^2 = -e^{2\alpha(r)} dt^2 + e^{2\beta(r)}dr^2 + r^2 d\Omega_N, \label{eq:StandardSSM_HD} 
\eeq
\noindent where 
\beq d \Omega_N = d\phi_1 d \phi_1 + \sum_{\mu = 2}^{N} S^2_{\mu} d \phi_\mu d \phi_\mu, \label{eq:Nsphere} \eeq
\noindent and 
\beq S_{\mu}(\phi_1, \ldots \phi_{\mu-1}) = \left( \prod_{\nu = 1}^{\mu-1}  \sin (\phi_\nu) \right). \eeq

There is an alternative coordinate system, $(u,r,\phi_1, \ldots \phi_N)$ for these metrics that is adapted to outgoing photons which is a generalization of Eddington-Finkelstein coordinates \cite{Wald:1984rg} known as outgoing coordinates by making the transformation
\beq 
    t=u+f(r), \quad f' = \frac12 e^{2\beta-2\alpha} \label{eq:EF_transf}
\eeq
\noindent Then denoting
\beq
    V=\frac{\sqrt{2}}{2} e^{\alpha-\beta},  e^{\gamma} = \frac{\sqrt{2}}{2}e^{\alpha + \beta} \label{eq:Sph_to_EF}
\eeq
\noindent the resulting metric becomes
\beq
    ds^2 = -2  e^{2\gamma} du (V du + dr) + r^2 d\Omega_N, \label{eq:NullSS_HD} 
\eeq
\noindent As a further benefit for this coordinate system, the expressions for the curvature tensor and its covariant derivatives relative to a null coframe basis are more compact than those in the spherical or Cartesian coordinate systems.

We will now briefly review the Cartan invariants for the static spherically symmetric spacetimes in the null coordinate system. Following this we will compute the Killing invariants for these spacetimes and compare them to the Cartan invariants. Finally we will consider the special case where the spacetimes admit a Killing horizon.

\subsection{Cartan invariants}

To compute the Cartan invariants, we will follow the natural modification of CK algorithm to higher dimensions by using the alignment classification \cite{milson2005alignment, ortaggio2012algebraic} in conjunction with the Segre classification \cite[Chapter 5]{kramer} to determine the necessary canonical forms. We will start with the following null coframe $\{ \theta^a \} = \{n,\ell, m^i \}$ in the coordinate system $(u,r,\phi_1, \ldots, \phi_N)$:

\beq 
\begin{aligned}
    \theta^1 = n &= e^{2\gamma} ( dr + V du), \\
    \theta^2 = \ell &= du, \\
    \theta^i = m^i &= r S_{i-2} d \phi_{i-2}
\end{aligned} \label{eq:Cartan_coframe}
\eeq

\noindent where $i,j,k \in \{ 3 \ldots N\}$ and we will define $S_1 = 1$. The non-vanishing connection coefficients are
\beq 
\begin{aligned}
    \Gamma_{121} &= - \Gamma_{211} = 2\gamma', \\
    \Gamma_{122} &= - \Gamma_{212} = e^{-2\gamma} V', \\
    \Gamma_{1ii} &= - \Gamma_{i1i} = - \frac{1}{r}, \\
    \Gamma_{2ii} &= - \Gamma_{i2i} =  \frac{e^{-2\gamma} V}{r}, \\
    \Gamma_{ijj} &= - \Gamma_{jij} = - \frac{\cos(\theta_{i-2})}{r S_{i-1}},~i<j. 
\end{aligned} \label{eq:SSS_Connect}
\eeq

\noindent The components of the curvature tensor are:
\beq
\begin{aligned}
    R_{1212} &= e^{-2\gamma} (2\gamma'' V + V'' + 2\gamma' V'), \\
    R_{1i1i} &= \frac{2\gamma'}{r}, \\
    R_{1i2i} &= \frac{e^{-2\gamma}V'}{r}, \\
    R_{2i2i} &= \frac{2\gamma' e^{-4\gamma} V^2}{r}, \\
    R_{ijij} &= \frac{1- 2e^{-2\gamma} V}{r^2},~i<j. 
\end{aligned} \label{eq:SSS_Riem}
\eeq
\noindent Due to spherical symmetry and the existence of a hypersurface orthogonal timelike Killing vector field, the number of functionally independent Cartan invariants produced from the CK algorithm will be at most one. Only in one particular case, discussed below will the number of functionally independent Cartan invariants be zero for static spherically symmetric spacetimes.

In principle, we could use the Weyl tensor and the Ricci tensor to further fix the frame. We will not explicitly show the Weyl tensor but we note that, in general, it is of aligned type {\bf D} or it vanishes and hence it is of alignment type {\bf O}. Employing the Segre classification there are three possibilities with the following conditions on $\gamma$ and $V$:
\begin{itemize}
    \item Case 1: $[1,1~(1,\ldots 1)]$, $\gamma' \neq 0$
    \item Case 2:  $[(1,1)~ (1,\ldots 1)]$, $\gamma' = 0$
    \item Case 3:  $[(1,1~1,\ldots 1)]$, $\gamma' = 0, V =\frac12 + m r^{3-N}+\Lambda r^2$.
\end{itemize}
\noindent The Ricci tensor can vanish in the third case when $m = \Lambda = 0$.

If the Weyl tensor is non-vanishing, the condition that it is of alignment type {\bf D} implies that the null vector fields $\ell$ and $n$ are determined up to a boost, i.e., $\ell' = \lambda \ell,~ n' = \lambda^{-1} n$. Thus, the linear isotropy group at zeroth order, $H_0$, is:
\begin{itemize}
    \item Case 1: $SO(N)$,
    \item Case 2 and Case 3: $SO(N)$ and boosts.
\end{itemize}

\noindent If the Weyl tensor vanishes, $H_0$ is then:
\begin{itemize}
    \item Case 1: $SO(N)$,
    \item Case 2: $SO(N)$ and boosts, 
    \item Case 3: $SO(1,N+1)$
\end{itemize}

This last case arises if and only if $m=0$. Furthermore, since $R_{abcd;e} = \nabla_e R_{abcd} = 0$, this case is either de Sitter or anti-de Sitter spacetime depending on the sign of $\Lambda$ or Minkowski spacetime if $\Lambda=0$. We will exclude this case from our analysis as all Cartan invariants consists of $\Lambda$. This solution is also the only solution that has no functionally independent invariants at zeroth order. We may assume the number of functionally independent invariants is always one at zeroth order and instead focus on the dimension of the linear isotropy group. 

Following from this discussion there are two cases that determine related to fixing of the boost parameter: whether $\gamma' = 0$. If $\gamma' \neq 0 $ then we may always apply a boost, $\lambda = e^{-\gamma} \sqrt{V}$ to set 
\beq 
    R_{1i1i} = R_{2i2i} = \frac{2\gamma'V e^{-2\gamma}}{r}. 
\eeq 
\noindent While if $\gamma' = 0$ the boost parameter may be fixed at first order. Due to the non-trivial isotropy group, the linear isotropy will be at most $SO(N-2)$ meaning that all subsequent covariant derivatives of the curvature tensor will be unchanged under spatial rotations of $m^i$.  

Relative to the coframe in equation \eqref{eq:Cartan_coframe}, the non-zero components of the covariant derivative of the curvature tensor are:
\beq 
\begin{aligned}
    R_{1212;1} &= -e^{-2\gamma}V R_{1212;2} = -e^{-2\gamma} (4(\gamma')^2 V' + 4 \gamma' \gamma ' V - 4 \gamma''V'- 2 \gamma'''V - V'''), \\
    R_{1i1i;1} &= -R_{2i2i;2} = -\frac{2e^{-2\gamma}(2 V'\gamma' r+ V \gamma'' r- \gamma' V)}{r}, \\
    R_{1i1i;2} &= - R_{2i2i;1} = - \frac{-2 e^{-2\gamma}(2 V' \gamma' r + V \gamma'' r -\gamma' V)}{r^2}, \\
    R_{ijij;1} &= -e^{-2\gamma}V R_{ijij;2} = - \frac{2(-2\gamma' e^{-2\gamma} Vr + e^{-2\gamma}V'r-2e^{-2\gamma}V +1)}{r^3}
\end{aligned} \label{eq:SSS_DRiem}
\eeq
\noindent If $\gamma' \neq 0$ then the components of $R_{abcd;e}$ may be transformed under a boost transformation with $\lambda = e^{-\gamma} \sqrt{V}$ to yield Cartan invariants that act as classifying functions for the manifold by employing the inverse function theorem, to rewrite this invariant in terms of some chosen functionally independent invariant.  

If $\gamma' = 0$, it is straightforward to show that if the boost parameter cannot be fixed at first order then the spacetime  is (anti)-de Sitter spacetime since $R_{abcd;e}= 0$. Thus, we may fix the boost parameter $\lambda$, for example, by setting $R_{1212;1} = R_{1212;2}$ or $R_{ijij;1} = R_{ijij;2}$. The resulting components are Cartan invariants that act as classifying functions for the spacetime. In this case, as the dimension of the linear isotropy group has changed, we must compute the second covariant derivative of the curvature tensor to recover all necessary Cartan invariants.  

To summarize the analysis of the CK algorithm for static spherically symmetric metrics with at most four Killing vector fields. If $\gamma' \neq 0$ the algorithm concludes at first order and the non-zero components $\mathcal{R} =\{ R_{abcd}, R_{abcd;e}$ relative to the boosted coframe such that $R_{1i1i} = R_{2i2i}$ constitute the set of all Cartan invariants. While if $\gamma' = 0$ then relative to the boosted coframe \eqref{eq:Cartan_coframe} where $R_{1212;1} = R_{1212;2}$ or $R_{ijij;1} = R_{ijij;2}$ the non-zero components of $\mathcal{R} = \{ R_{abcd}, R_{abcd;e}, R_{abcd;f}\}$ yield the necessary Cartan invariants. Given two metrics, $g$ and $g'$ that share the same canoncial forms for the curvature tensor and its covariant derivatives, then equating $\mathcal{R} = \mathcal{R}'$ as ordered lists yield the equations for which a solution must be found to determine equivalence.

\subsection{Killing invariants}

The Lie algebra of Killing vector fields is isomorphic to $\mathbb{R} \times \mathfrak{so}(N+1)$. In Cartesian coordinates, the vector fields lying in $\mathfrak{so}(N+1)$ vanish at some point and hence lie in the isotropy group. However, in these coordinates, we have two translational Killing vector fields,
\beq {\bf X}_1 = \partial_t, \quad\quad {\bf X}_2 = \partial_{\theta_N}, \eeq
\noindent or as one-forms: 
\beq {\bf F}^1 = e^{2\alpha} dt, \quad\quad{\bf F}^{2} = r^2S_N^2 d \phi_N. \eeq

\noindent The second one-form is well-defined except when $\theta_N =0$. With these vector fields, we may choose them as frame elements almost everywhere in the geometry.

In spherical coordinates, the Gram-Schmidt procedure yields the following basis written as a null coframe, $\{\theta^a\} = \{ n,\ell, \bar{m}, m, m_\mu\}$, \cite{kramer,garcia2009spinor}
\beq \begin{aligned}
n & = \frac{1}{\sqrt{2}}( e^\alpha dt - r S_N d\phi_N), \\[10pt]
\ell & = \frac{1}{\sqrt{2}}( e^\alpha dt + r S_N d\phi_N), \\[10pt]
\bar{m} & = \frac{1}{\sqrt{2}}( e^\beta dr + i r d \phi_1), \\[10pt]
m &= \frac{1}{\sqrt{2}}( e^\beta dr - i r d\phi_1 ), \\[10pt]
m_\mu &= r S_\mu d \phi_\mu, \quad\quad \mu = 2,\ldots, N-1 .
\end{aligned} \eeq

\noindent At zeroth-order, we find two non-trivial Killing invariants, 
\beq N_{11} = e^{2\alpha}, \quad\quad N_{22} = r S_N .\eeq

Taking the exterior derivatives of ${\bf F}^1$ and ${\bf F}^2$, we find
\beq \begin{aligned} d {\bf F}^1 &= -4 \alpha_{,r} e^{\alpha-\beta} (m + \bar{m}) \wedge (n+\ell), \\
d {\bf F}^2 &= -e^{-\beta} S_N (m+\bar{m}) \wedge (n-\ell ) - i S_N \frac{\cos(\phi_1)}{\sin(\phi_1)}  (m - \bar{m}) \wedge (n - \ell) \\
&~~~- \sum_{\mu = 2}^{N-1} \frac{\sqrt{2} S_N }{S_{\mu}} \frac{\cos(\phi_\mu)}{\sin(\phi_\mu)} [m^{\mu} \wedge (\ell-n)].  \end{aligned} \eeq

\noindent The remaining parameters for the boost and spatial rotations can be set to identity, as the components of the two-forms are described entirely by their squared-norms. To conclude the algorithm, we would compute the exterior derivatives of the components $\{ F^I_{~ab}\}$, as the linear isotropy group would remain trivial and indicate the algorithm stops at order $2$. Thus $\mathcal{K}_2$ contains all relevant information and the static spherically symmetric metrics can be classified using the classifying functions arising from this set. 


\subsection{Static spherically symmetric metrics with a Killing horizon}

A Killing horizon is defined as a null hypersurface where a Killing vector becomes null and acts a generator for the null hypersurface \cite{Wald:1984rg}. On such a surface, it is no longer possible to build an orthonormal frame adapted to all of the translational Killing vector fields since at least one of them is now a null generator for the hypersurface. Despite this, Killing invariants can be used in any region exterior to this surface. 

In any dimension, such a metric takes the following form in advanced coordinates
\beq ds^2 = - e^{2 \gamma(r)} \left( 1 - \frac{2m(r)}{r}\right) dv^2 +2e^{ \gamma(r)} dvdr + r^2 d \Omega_N, \eeq

\noindent where $d\Omega_N$ is defined in Equation \eqref{eq:Nsphere}. In this coordinate system the timelike Killing vector is given as $ {\bf X}_1 = \partial_v$. Using the transformation $t = v + f(r)$, where the function $f$ satisfies $f'(r) = -\frac12 e^{-\gamma(r)} \left(1-\frac{2m(r)}{r}\right)^{-1}$, the metric becomes 
\beq ds^2 = - e^{2\gamma(r)} \left( 1 - \frac{2 m(r)}{r}\right) dt^2 + \left( 1 - \frac{2 m(r) }{r}\right)^{-1} dr^2 + r^2 d\Omega_N. \eeq

\noindent Comparing with the general metric in Equation \eqref{eq:StandardSSM_HD}, it follows that 
\beq \alpha(r) = \gamma(r) + \frac12 \ln \left( 1 - \frac{2m(r)}{r}\right), \quad\quad \beta(r) = - \frac12 \ln \left( 1 - \frac{2m(r)}{r}\right), \eeq
\noindent while two relevant Killing invariants take the form
\beq \begin{aligned} N_{11} &= e^{2 \gamma(r)} \left( 1 - \frac{2m(r)}{r}\right), \\
F^1_{~13} &= \frac{e^{\gamma(r)}}{r^2} \qty[ r \gamma(r)_{,r}  (2 m(r) - r) + rm(r)_{,r} - m(r)]. \end{aligned} \eeq

Given a scalar invariant, its zero-set specifies an invariantly-defined hypersurface, which can have physical significance for the spacetime. In the current expressions we may determine coordinate values for the Killing horizon as their zero-set. In general, the second invariant, $F^1_{~13}$, will not vanish on the horizon, $r=2m$, except for special choices of $m(r)$; thus, we can choose $F^1_{~13}$ as the functionally independent invariant. Using the inverse function theorem, we may express the first invariant in terms of the other, which we will call a classifying curve for the space as this expression will be independent of coordinates:
\beq I_{r} \equiv F^1_{~13}; \quad\quad N_{11} = e^{2\gamma'(I_r)} \left( 1 - \frac{2m'(I_r)}{r(I_r)} \right). \eeq 
\noindent If both of the invariants vanish on the Killing horizon, due to the choice of $m(r)$, the ratio of these two Killing invariants is again a Killing invariant, which is non-zero at all points in the manifold. We may then write down a different classifying curve for the space:
\beq I_r \equiv \frac{N_{11}}{F^1_{~13}};\quad\quad  N_{11} = \sqrt{I_r}\left( 1 - \frac{2m'(I_r)}{r(I_r)} \right), \quad\quad F^1_{~13} = \frac{1}{\sqrt{I_r}} \left( 1 - \frac{2m'(I_r)}{r(I_r)} \right). \eeq

\noindent Thus the Killing horizon can be expressed in an invariant manner, independent of any coordinate system. Furthermore, this can be generalized to solutions that may contain multiple Killing horizons, such as Schwarzschild(-anti)-de Sitter solutions or Reissner-Nordstr{\"o}m solutions.  

\section{Discussion}

In this paper, we have introduced a new formalism and set of invariants for classifying spacetimes admitting isometries. Any isometry of a space is generated by a Killing vector field, which is necessarily an invariant vector field of the spacetime and can be found by solving the Killing equations. Furthermore, the norm of any vector field is an invariant quantity which is independent of the coordinate system chosen. By adapting a frame to the set of translational Killing vector fields, it is possible to work with their norms and exterior derivatives as an exterior differential system and generate a finite list of invariants. 

While determination of an invariant frame is a common tool in both the Cartan-Karlhede (CK) algorithm \cite{kramer} and Cartan's method of moving frames \cite{olver1995equivalence}, in general this approach is distinct from them. Only in the case of locally homogeneous or maximally symmetric spaces does this approach reduce to the Cartan's method of moving frames where the translational Killing vector fields yield the necessary coframe basis. However, our algorithm will never specialize to the CK algorithm since it does not use the irreducible parts of the curvature tensor or their covariant derivatives. Furthermore, the linear isotropy group of the CK algorithm will generically be larger than the linear isotropy group arising in our algorithm. 

In future work, we will employ the algorithm to characterize axisymmetric and stationary solutions to Einstein's field equations and other modified gravity theories in order to characterize these solutions and identify the subset of spacetimes which admit Boyer-Lindquist coordinates. Of particular interest are deformations of Kerr-like solutions \cite{pugliese2022wormholes} and whether such deformations preserve the required irreducible rank 2 Killing tensor \cite{eichhorn2021locality}. 

\section*{Acknowledgments}
\noindent The authors would like to thank Eivind Schneider for their discussion and helpful comments. DM is supported by the Norwegian Financial Mechanism 2014-2021 (project registration number 2019/34/H/ST1/00636).

\bibliographystyle{unsrt-phys-eucos}
\bibliography{ref}

\end{document}